\documentclass[aps,prl,twocolumn,superscriptaddress,showpacs]{revtex4}

\usepackage{graphicx}
\usepackage{dcolumn} 
\usepackage{bm}      

\begin{document}

\title{Competition between the structural phase transition and superconductivity in Ir$_{1-x}$Pt$_x$Te$_2$ as revealed by pressure effects}

\author{A.~Kiswandhi}
\affiliation{National High Magnetic Field Laboratory, Florida State University, Tallahassee, FL 32306-4005, USA}
\affiliation{Department of Physics, Florida State University, Tallahassee, FL 32306-3016, USA}

\author{J.~S.~Brooks}
\affiliation{National High Magnetic Field Laboratory, Florida State University, Tallahassee, FL 32306-4005, USA}
\affiliation{Department of Physics, Florida State University, Tallahassee, FL 32306-3016, USA}

\author{H.~B.~Cao}
\affiliation{Quantum Condensed Matter Division, Oak Ridge National Laboratory, Oak Ridge, Tennessee 37831, USA }

\author{J.~Q.~Yan}
\affiliation{Materials Science and Technology Division, Oak Ridge National Laboratory, Oak Ridge, Tennessee 37831, USA }
\affiliation{Department of Materials and Engineering, University of Tennessee, Knoxville, Tennessee 37996, USA}

\author{D.~Mandrus}
\affiliation{Materials Science and Technology Division, Oak Ridge National Laboratory, Oak Ridge, Tennessee 37831, USA }
\affiliation{Department of Materials and Engineering, University of Tennessee, Knoxville, Tennessee 37996, USA}

\author{Z.~Jiang}
\affiliation{School of Physics, Georgia Institute of Technology, Atlanta, Georgia 30332, USA}

\author{H.~D.~Zhou}
\email{hzhou10@utk.edu}
\affiliation{National High Magnetic Field Laboratory, Florida State University, Tallahassee, FL 32306-4005, USA}
\affiliation{Department of Physics and Astronomy, University of Tennessee, Knoxville, Tennessee 37996-1200, USA}

\date{\today}

\begin{abstract}
Pressure-dependent transport measurements of Ir$_{1-x}$Pt$_x$Te$_2$ are reported. With increasing pressure, the structural phase transition at high temperatures is enhanced while its superconducting transition at low temperatures is suppressed. These pressure effects make Ir$_{1-x}$Pt$_x$Te$_2$ distinct from other studied $T$X$_2$ systems exhibiting a charge density wave (CDW) state, in which pressure usually suppresses the CDW state and enhances the superconducting state. The results reveal that the emergence of superconductivity competes with the stabilization of the low-temperature monoclinic phase in Ir$_{1-x}$Pt$_x$Te$_2$.
\end{abstract}


\maketitle

One of the fundamental interests in the physics of transition metal dichalcogenides ($T$X$_2$, $T$ = Ti, Ta, or Nb, $X$ = S, Se, or Te) with 1$T$ and 2H structures is the competition between two very different phenomena: a charge density wave (CDW) state and superconductivity \cite{CDW1, CDW2, CDW3}.  Although the studies on $T$X$_2$ have been ongoing for almost half  a century, the microscopic origin of the CDW state is not yet fully understood. On the other hand, studies of $T$X$_2$ systems have revealed interesting physical behavior \cite{ CDW4,TiSepressure, CuTiSe, TaSpressure, NbSe}.

\begin{figure}[tbp]
\linespread{1}
\par
\begin{center}
\includegraphics[width=3.2in]{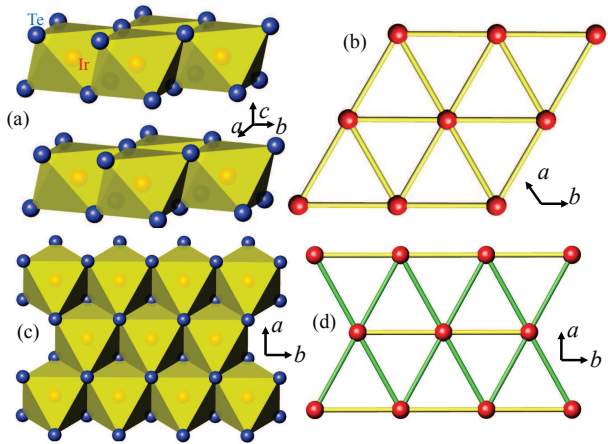}
\end{center}
\par
\caption{(Color Online) (a) The lattice structure of IrTe$_2$ in the trigonal phase at $T$ $>$ $T_s$ showing (b) the equilateral triangular lattice of Ir atoms in the $ab$ plane. (c) The lattice structure of IrTe$_2$ in the monoclinic phase at $T$ $<$ $T_s$ showing (d) the isosceles triangular lattice of Ir atoms in the $ab$ plane. In (d) the two green lines represent the short Ir-Ir bonds.}
\end{figure}

Recently another member of the $T$X$_2$ family with a 5$d$ transition metal, IrTe$_2$, has received a lot of attention due to its properties: (I) At the room temperature, the 1$T$-IrTe$_2$ crystallizes in a trigonal structure with the edge-sharing IrTe$_6$ octahedra forming layers stacked along the $c$ axis [Fig. 1(a)] with the Ir ions forming equilateral triangular lattices in the $ab$ plane [Fig. 1(b)]. By decreasing temperature, at around 250 K shorter Ir-Ir bonds form and the strucure transforms into a monoclinic structure [Fig. 1(c, d)]. This transition is accompanied with a jump in resistivity and a decrease in susceptibility \cite{IrTeres}. This phenomenon is similar to that of the CDW state in other $T$X$_2$ systems. However the early NMR studies did not support the CDW transition scenario in IrTe$_2$ \cite{IrTeNMR}. Recent electron diffraction \cite{PdIrTe} and photoemission studies \cite{PhotoIrTe} both suggested that the Ir $t_{2g}$ orbitals could contribute in the transition. Therefore, the exact origin of this transition is still unclear. (II) Due to its large atomic number, IrTe$_2$ is expected to show a strong spin-orbital (SO) coupling, which has been shown to result in unique quantum states in other materials, e.g. the $J_{\text{eff}}$ = 1/2 Mott insulator in Sr$_2$IrO$_4$ \cite{SrIrO} and topological insulators \cite{TI1, TI2, TI3}. (III) More interestingly, recent studies show that a small percentage of Pd or Pt doping induces superconductivity at temperatures below 4 K in the parent compound \cite{PdIrTe, PtIrTe}. However, it is still unclear whether the origin of the superconducting transition is similar to that of the other $T$X$_2$ systems with a CDW state.

In this Rapid Communication, we use hydrostatic pressure as a tool to influence the transitions in Ir$_{1-x}$Pt$_x$Te$_2$. Transport measurements performed at different pressures clearly show that the structural phase transition temperature increases while the superconducting transition temperature decreases with increasing pressure. These pressure effects make IrTe$_2$ distinct from most other $T$X$_2$ systems in which pressure usually suppresses the CDW states while enhancing the superconductivity. The results show that Ir$_{1-x}$Pt$_x$Te$_2$ is a unique system due to the competition between the structural and superconducting transitions.

Single crystals of IrTe$_2$ and Ir$_{0.95}$Pt$_{0.05}$Te$_2$ were grown using a self-flux technique as recently reported \cite{IrTeSC}. The polycrystalline samples Ir$_{0.98}$Pt$_{0.02}$Te$_2$ and  Ir$_{0.97}$Pt$_{0.03}$Te$_2$ were prepared by standard solid state reaction. To avoid the grain boundary effect in transport measurements, the dense and hard pellets of the polycrystalline samples were prepared by a cold-press technique \cite{CP}. The transport measurements were performed using a four probe technique and the hydrostatic pressure was generated by using Daphne oil 7373 as a pressure medium in a standard piston cylinder clamp cell. At low temperature, the pressure generated by the Daphne oil 7373 is known to decrease by approximately 1.5 kbar \cite{Daphne7373}. Any pressure mentioned in this paper will be the pressure as measured at the room temperature. Single-crystal neutron diffraction measurements were performed at the HB-3A four-circle diffractometer at the High Flux Isotope Reactor at the Oak Ridge National Laboratory.

\begin{figure}[tbp]
\linespread{1}
\par
\begin{center}
\includegraphics[width=3.2 in]{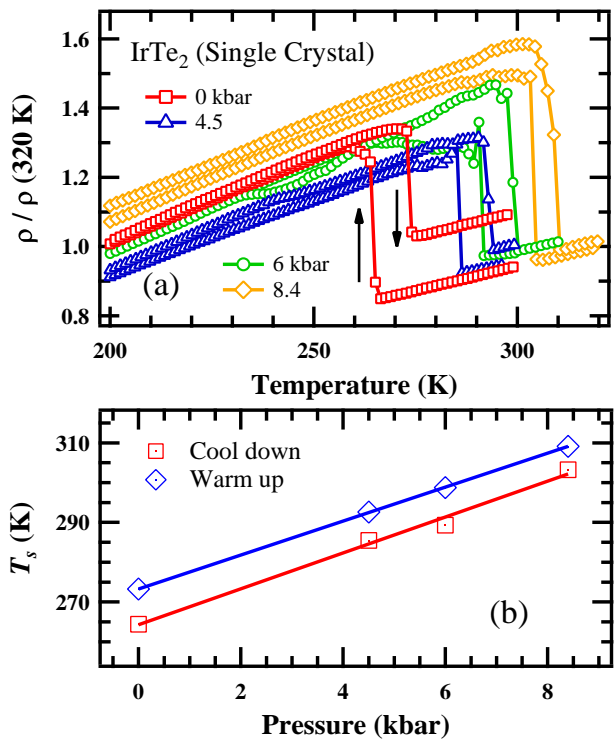}
\end{center}
\par
\caption{(Color Online) (a) Temperature dependence of resistivity of IrTe$_2$ at various pressures showing a clockwise thermal hysteresis around the structural transition. For P $<$ 8.4 kbar, $\rho$(320 K) was obtained by extrapolating the linear part at $T$ $>$ $T_s$; (b) Pressure dependence of $T_s$ for cool-down and warm-up runs.}
\end{figure}

The structural phase transition temperature was determined from the transport measurement since the transition is accompanied with an increase in the resistivity ($\rho$). Here, $T_s$ is defined as the minimum of d$\rho$/d$T$.  At the ambient pressure $T_s$ was found to be 264 K and 272 K for IrTe$_2$ during the cool down and warm up process, respectively [Fig. 2(a)], which is consistent with a previous report \cite{IrTeSC}. With increasing pressure, $T_s$ increases linearly with d$T_s$/d$P$ = 4.393 K/kbar [Fig. 2(b)]. Here d$T_s$/d$P$ is the average of d$T_s$/d$P$ during the cool-down and warm-up process, as the values are very similar. The transport measurement shows a thermal hysteresis, characteristic of a first-order transition. IrTe$_2$ shows no superconductivity down to 1.3 K at all measured pressures (not shown).

\begin{figure}[tbp]
\linespread{1}
\par
\begin{center}
\includegraphics[width=3.2 in]{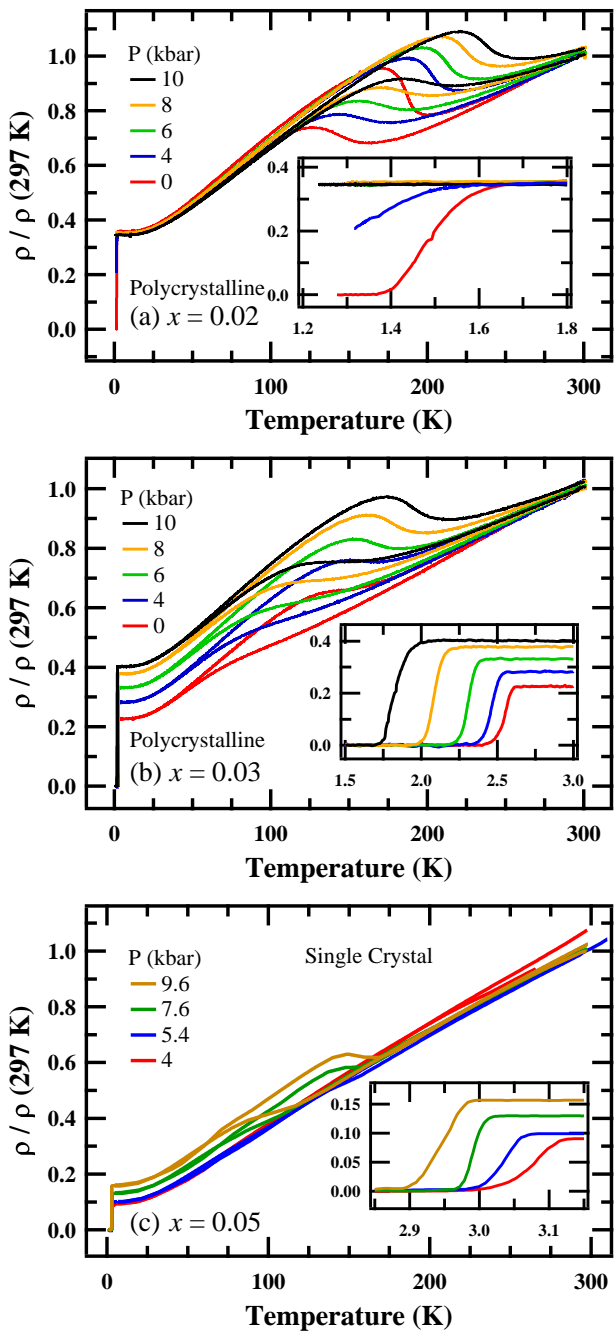}
\end{center}
\par
\caption{(Color Online) Temperature dependence of normalized resistivity for Ir$_{1-x}$Pt$_x$Te$_2$ at various pressures. Polycrystalline samples with (a) $x$ = 0.02 and (b) $x$ = 0.03.  (c) Single crystal with $x$ = 0.05. The insets show the superconducting transition part of each corresponding figure.}
\end{figure}

In Ir$_{1-x}$Pt$_x$Te$_2$ samples at ambient pressure $T_s$ is suppressed to lower temperatures, and superconductivity appears below 4 K with increasing Pt doping (Fig. 3), which is consistent with the reported data \cite{PtIrTe}. For $x$ = 0.02 and 0.03, $T_s$ increases linearly with increasing pressure and the rates are 5.235 K/kbar and 4.098 K/kbar, respectively. For $x$ = 0.03, the superconducting transition temperature, $T_{\text{SC}}$, shifts toward lower temperatures with increasing pressure with a rate d$T_{\text{SC}}$/d$P$ = - 0.09 K/kbar [inset of Fig. 3(b)]. For $x$ = 0.02, the superconducting transition is not yet complete at P $\geq$ 4 kbar at the lowest accessible temperature. The single crystal $x$ = 0.05 shows no detectable anomaly or hysteresis as apparent as in $x \leq$ 0.03, at P $\leq$ 5.4 kbar. Similar to other samples, its $T_{\text{SC}}$ decreases with increasing pressure with a rate of  - 0.025 K/kbar. However, at 7.6 kbar, the thermal hysteresis reappeared at $T_s$ = 96.5 K and 151.6 K during cool-down and warm-up, respectively, for $x$ = 0.05. At 9.6 kbar $T_s$ increases to 100.7 K and 156 K, respectively.

For $x$ = 0, 0.02, and 0.03, it is clear that the resistivity jump is related to the structural phase transition. The increase of $T_s$ for these three samples with increasing pressure indicates the low-temperature monoclinic phase is stabilized by pressure. The Pt doping completely suppresses the structural phase transition for $x$ = 0.05 at low pressures and as a result no resistivity jump or hysteresis were observed. The reappearance of the resistivity anomaly for $x$ = 0.05 at high pressures, with features similar to that of the low doping samples, clearly indicates that the structural phase transition is recovered in this sample at high pressures above 7.6 kbar. It is also noticed that for $x$ = 0.02, the resistivity increases around $T_s$ with increasing pressure; and for $x$ = 0.03 and 0.05, the resistivity increases for the whole temperature region below $T_s$ with increasing pressure.

\begin{figure}[tbp]
\linespread{1}
\par
\begin{center}
\includegraphics[width=3.2in]{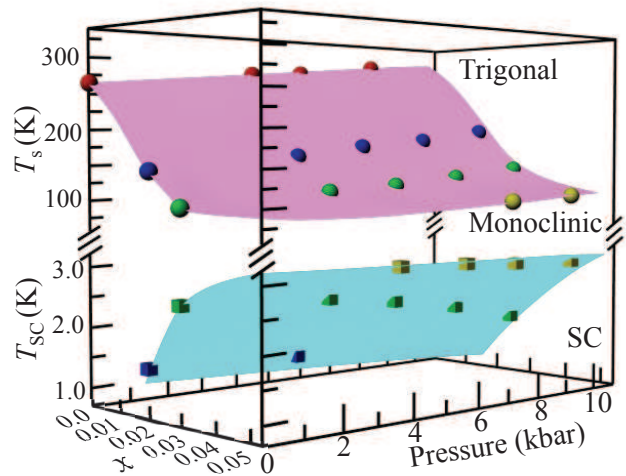}
\end{center}
\par
\caption{(Color Online) The phase diagram of Ir$_{1-x}$Pt$_x$Te$_2$. The spheres and the cubes represent the experimental values of $T_s$ and $T_{\text{SC}}$, respectively. Here the $T_s$ values are obtained from the cool-down process. The surfaces are the schematic phase boundaries separating the trigonal (metallic), monoclinic (metallic), and superconducting phases.}
\end{figure}

The overall behavior of the system revealed by pressure effects are summarized in Fig. 4. Using pressure as controlled parameter, the structural transition temperature of Ir$_{1-x}$Pt$_x$Te$_2$ is enhanced in temperatures and the superconducting transition temperature is reduced. For $x$ = 0.05, the structural phase transition, suppressed by Pt-doping, is recovered by high pressure. The competition between the structural phase transition and superconductivity is clearly revealed.

In some of the $T$X$_2$ systems, such as 2H-TaSe$_2$, 2H-TaS$_2$, and 2H-NbSe$_2$, the CDW transition coexists with the superconductivity \cite{CDW1, CDW2, CDW3, CDW4}. On the other hand, in 1$T$-TiSe$_2$ \cite{TiSepressure, CuTiSe} and 1$T$-TaS$_2$ \cite{TaSpressure, NaTaS}, the superconductivity can be achieved by doping or pressure. Generally, for $T$X$_2$ systems, the doping mainly shifts the chemical potential into the conduction band due to donor electrons and the pressure mainly increases $N$($E_F$) by suppressing the CDW state and restoring the Fermi surface to the undistorted state. Therefore, both doping and pressure have similar effects on $T$X$_2$ systems: suppressing the CDW transition and enhancing the superconductivity \cite{TiSepressure, CuTiSe, TaSpressure, Pressure1, Pressure2, Pressure3, Pressure4, Pressure5}. The pressure studies on $T$X$_2$ without introducing any degree of internal disorder by doping clearly demonstrates the competition between the CDW state and superconductivity.

It is apparent that the pressure effects on 1$T$-IrTe$_2$ and related Pt-doped samples are different from other $T$X$_2$ systems. The reported powder x-ray diffraction studies on IrTe$_2$ show that the high temperature transition is of a structural nature from a trigonal to a monoclinic structure with space group $C$2/$m$. The transition causes the Ir-Ir distance to contract along the $b$ axis, establishing an Ir-Ir bond. On the other hand, the distance between the IrTe$_6$ layers along the $c$ axis expands \cite{IrTeres, PtIrTe}. Based on the similarity of the structural distortion between IrTe$_2$ and another triangular lattice material NaTiO$_2$ with Ti$^{3+}$ $t_{2g}$ orbital ordering \cite{NaTiO1, NaTiO2}, Pyon $et~al$ \cite{PtIrTe} proposed that the structural phase transition in IrTe$_2$ could be induced by the ordering of the Ir 5$d$ $t_{2g}$ orbitals. The most recent photoemission results also suggest an orbitally induced Peierls effect in IrTe$_2$ \cite{PhotoIrTe}. Moreover, Yang $et~al$ \cite{PdIrTe} proposed a charge-orbital density wave scenario for IrTe$_2$, based on the fact that the structural phase transition is accompanied by the emergence of superlattice peaks. Although the exact origin of the structural phase transition for IrTe$_2$ is unclear, all of the studies indicate that the transition is not exactly CDW based as shown by other $T$X$_2$ systems but more likely related to orbital ordering. Another fact is that no electronic phase transition is induced by the structural phase transition in IrTe$_2$, but only a resistivity jump in a generally metallic behavior over the entire temperature region.  Therefore, in IrTe$_2$ the structural distortion is the main phenomenon leading to the observed transport behavior. For other $T$X$_2$ systems with a CDW transition, the lattice deformation is usually regarded as being caused by electronic ordering. This fundamental difference could be the primary reason for the different pressure effects between IrTe$_2$ and other $T$X$_2$ systems.

Recently, single crystal x-ray and neutron diffraction measurements on an as-prepared IrTe$_2$ sample were also performed to revisit its structure \cite{structure}. The results confirmed that the low-temperature phase is a monoclinic phase, but with the contraction of two of the three nearest Ir-Ir bonds instead of only one as in the previously reported powder x-ray diffraction studies. An orbital ordering transition of the Ir sites is proposed to account for these two bond contractions and the resulting structural phase transition with the IrTe$_6$ octahedra distortion.  More importantly, the volume shows a 1.0\% decrease below the structural phase transition \cite{structure}. Therefore, when the applied pressure contracts the IrTe$_2$  volume, the low-temperature, smaller volume monoclinic phase is stabilized, resulting in an increase in $T_s$. The stabilization of the low temperature phase by pressure is also shown by the increase of the resistivity below $T_s$ with increasing pressure for Pt-doped samples, since the low-temperature monoclinic phase favors higher resistivity due to the resistivity jump around $T_s$. The pressure effects on IrTe$_2$ are analogous to those of spinel CuIr$_2$S$_4$. In CuIr$_2$S$_4$, the structural transition occurs from cubic to tetragonal with a volume contraction of 0.7\% \cite{CuIrS1} due to the orbitally induced Peierls state below $T_s$ \cite{CuIrS2}. The transition is accompanied by a metal-insulator transition due to the charge ordering of Ir$^{3+}$-Ir$^{4+}$ electrons \cite{CuIrS3}. In contrast to most systems exhibiting metal-insulator transitions, high pressure stabilizes the low-temperature insulating phase for CuIr$_2$S$_4$, leading to an increase in $T_s$ and resistivity below $T_s$ \cite{CuIrS4}. While the d$T_s$/d$P$ for CuIr$_2$S$_4$ was found to be about 2.8 K/kbar, smaller than that of Ir$_{1-x}$Pt$_x$Te$_2$, which is 4.0 $\sim$ 5.0 K/kbar, the similarity of the phenomenon could be useful to understand the exact origin of the structural phase transition in IrTe$_2$ and more studies are needed.

In the Pt-doped samples, while the monoclinic phase is stabilized by pressure, the superconductivity is suppressed simultaneously. In the case of $x$ = 0.05, the structural phase transition is recovered with high pressures at 7.6 kbar. These pressure effects clearly show the competition between the structural phase transition and superconductivity: the stabilization of monoclinic phase suppresses the superconductivity. On the other hand, with increasing Pt doping, $T_s$ is suppressed and superconductivity emerges. Pyon $et~al$ have already pointed out that the breaking of Ir-Ir bonds in the monoclinic phase by Pt doping results in the appearance of superconductivity.  The pressure and Pt doping affect the transitions in the opposite way. However, both effects consistently show that the structural fluctuations related to Ir-Ir bond formation in Ir$_{1-x}$Pt$_x$Te$_2$ are critical for the emergence of superconductivity. While the Ir-Ir bond in the monoclinic phase is stabilized by pressure (is broken by doping), the superconductivity is suppressed (is induced by doping).

In conclusion, the transport measurement at different pressures on Ir$_{1-x}$Pt$_x$Te$_2$ clearly revealed the competition between the structural phase transition at high temperatures and the superconductivity at low temperatures. This makes Ir$_{1-x}$Pt$_x$Te$_2$ unique compared with most $T$X$_2$ systems that have been studied, in which the competition is between two phases: the CDW state and superconductivity. Ir$_{1-x}$Pt$_x$Te$_2$ provides a unique case to study the close relationship between structural fluctuations and superconductivity.

\begin{acknowledgments}
The authors would like to thank Stanley Tozer, Vaughan Williams, Daniel McIntosh, Robert Schwartz, and David Graf for technical assistance concerning the pressure cell used in this work. This work is supported by NSF-DMR-0654118 and the State of Florida. A.K. is supported in part by NSF-DMR-1005293. Work at ORNL was supported by the US Department of Energy, Office of Basic Energy Sciences, the Scientific User Facilities Division (H.B.C), and the Materials Science and Engineering Division (J.Q.Y. and D.G.M.).

\end{acknowledgments}


\begin{thebibliography}{99}
\bibitem{CDW1} R.~L.~Withers and J.~A.~Wilson, J. Phys. C {\bf 19}, 4809  (1986).
\bibitem{CDW2} J.~A.~Wilson and A.~D.~Yoffe, Adv. Phys. {\bf 18}, 193  (1969).
\bibitem{CDW3}  J.~A.~Wilson, F.~J.~DiSalvo, and S.~Mahajan, Adv. Phys. {\bf 24}, 117  (1975).
\bibitem{CDW4} A.~H.~Castro Neto, Phys. Rev. Lett. {\bf 86}, 4382 (2001).
\bibitem{TiSepressure} A.~F.~Kusmartseva, B.~Sipos, H.~Berger, L.~Forr\'{o}, and E.~Tuti\v{s}, Phys. Rev. Lett. {\bf 103}, 236401 (2009).
\bibitem{CuTiSe} E.~Morosan, H.~W.~Zandbergen, B.~S.~Dennis, J.~W.~G.~Bos, Y.~Onose, T.~Klimczuk, A.~P.~Ramirez, N.~P.~Ong, and R.~J.~Cava, Nature Phys. {\bf 2}, 544 (2006).
\bibitem{TaSpressure} B.~Sipos, A.~F.~Kusmartseva, A.~Akrap, H.~Berger, L.~Forr\'{o}, and E.~Tuti\v{s}, Nature Mater. {\bf 7}, 960  (2008).
\bibitem{NbSe} T.~Valla, A.~V.~Fedorov, P.~D.~Johnson, P-A Glans, C. McGuinness, K.~E.~Smith, E.~Y.~Andrei, and H.~Berger, Phys. Rev. Lett. {\bf 92}, 086401, (2004).
\bibitem{IrTeres} N.~Matsumoto, K.~Taniguchi, R.~Endoh, H.~Takano, and S.~Nagata, J. Low. Temp. Phys. {\bf 117}, 1129 (1999).
\bibitem{IrTeNMR} K.~Mizuno, K.~Magishi, Y.~Shinonome, T.~Saito, K.~Koyama, N.~Matsumoto, and S.~Nagata, Physica B {\bf 312-313}, 818 (2002).
\bibitem{PdIrTe} J.~J.~Yang, Y.~J.~Choi, Y.~S.~Oh, A.~Hogan, Y.~Horibe, K.~Kim, B.~I.~Min, and S-W.~Cheong, Phys. Rev. Lett. {\bf 108}, 116402  (2012).
\bibitem{PhotoIrTe} D.~Ootsuki, Y.~Wakisaka, S.~Pyon, K.~Kudo, M.~Nohara, M.~Arita, H.~Anzai, H.~Namatame, M.~Taniguchi, N.~L.~Saini, and T.~Mizokawa, Phys. Rev. B {\bf 86}, 014519  (2012).
\bibitem{SrIrO} B.~J.~Kim, H.~Jin, S.~J.~Moon, J.~-Y.~Kim, B.~-G.~Park, C.~S.~Leem, J.~Yu, T.~W.~Noh, C.~Kim, S.~-J.~Oh, J.~-H.~Park, V.~Durairaj, G.~Cao, and E.~Rotenberg, Phys. Rev. Lett. {\bf 101}, 076402 (2008).
\bibitem{TI1} Y.~Xia, D.~Qian, D.~Hsieh, L.~Wray, A.~Pal, H.~Lin, A.~Bansil, D.~Grauer, Y.~S.~Hor, R.~J.~Cava, and M.~Z.~Hasan, Nature Phys. {\bf 5}, 398 (2009).
\bibitem{TI2} M.~Z.~Hasan and C.~L.~Kane, Rev. Mod. Phys. {\bf 82}, 3045 (2010).
\bibitem{TI3} X.~-L.~Qi and S.~-C.~Zhang, Rev. Mod. Phys. {\bf 83}, 1057 (2011).
\bibitem{PtIrTe} S.~Pyon, K.~Kudo, and M.~Nohara, J. Phys. Soc. Jpn. {\bf 81}, 053701 (2012).
\bibitem{IrTeSC} A.~F.~Fang, G.~Xu, T.~Dong, P.~Zheng, and N.~L.~Wang, Scientific Report, 1153/DOI:10.1038/sep01153.
\bibitem{CP} J.~-S. Zhou, J.~B.~Goodenough, and B.~Dabrowski, Phys. Rev. B {\bf 67}, 020404(R) (2003).
\bibitem{Daphne7373}K. Murata, H.~Yoshino, H.~O.~Yadav, Y.~Honda, and N.~Shirakawa, Rev. Sci. Instrum. \textbf{68}, 2490 (1997)
\bibitem{NaTaS} L.~Fang, Y.~Wang, P.~Y.~Zou, L.~Tang, Z.~Xu, H.~Chen, C.~Dong, L.~Shan, and H.~H.~Wen, Phys. Rev. B {\bf 72}, 014534 (2005).
\bibitem{Pressure1} T.~F.~Smith, L.~E.~Delong, A.~R.~Moodenbaugh, T.~H.~Geballe, and R.~E.~Schwall, J. Phys. C {\bf 5}, L230 (1972).
\bibitem{Pressure2} A.~H.~Thompson, F.~R.~Gamble, and R.~F.~Koehler, Phys. Rev. B {\bf 5}, 2811 (1972).
\bibitem{Pressure3} P.~Molini\'{e}, D.~J\'{e}rome, and A.~J.~Grant, Philos. Mag. {\bf 30}, 1091 (1974).
\bibitem{Pressure4} C.~Berthier, P.~Molini\'{e}, and D.~J\'{e}rome, Solid State Commun. {\bf 18}, 1393 (1976).
\bibitem{Pressure5} R.~H.~Friend and D.~J\'{e}rome, J. Phys. C {\bf 12}, 1441 (1979).
\bibitem{NaTiO1} K.~Takeda, K.~Miyake, K.~Takeda, and K.~Hirakawa, J. Phys. Soc. Jpn. {\bf 61}, 2156 (1992).
\bibitem{NaTiO2} S.~J.~Clarke, A.~J.~Fowkes, A.~Harrison, R.~M.~Ibberson, and M.~J.~Rosseinsky, Chem. Mater. {\bf 10}, 372 (1998).
\bibitem{structure} H.~B.~Cao, B.~C.~Chakoumakos, J.~Q.~Yan, H.~D.~Zhou, R.~Custelcean, and D.~Mandrus, arXiv:1302.5369v1.
\bibitem{CuIrS1} T.~Furubayashi, T.~Matsumoto, T.~Hagino, and S.~Nagata, J. Phys. Soc. Jpn. {\bf 63}, 3333 (1994).
\bibitem{CuIrS2} D.~I.~Khomskii and T.~Mizokawa, Phys. Rev. Lett. {\bf 94}, 156402 (2005).
\bibitem{CuIrS3} P.~G.~Radaelli, Y.~Horibe, M.~J.~Guttman, H.~Ishibashi, C.~H.~Chen, R.~M.~Ibberson, Y.~Koyama, Y.~-S.~Hor, V.~Kiryukhin, and S.~-W.~Cheong, Nature {\bf 416}, 155 (2002).
\bibitem{CuIrS4} G.~Oomi, T.~Kagayama, I.~Yoshida, T.~Hagino, and S.~Nagata, J. Magn. Magn. Mater. {\bf 140-144}, 157 (1995).
\end{thebibliography}
\end{document}